\documentclass[twocolumn]{revtex4}
\usepackage{amsmath}
\usepackage{amssymb}
\usepackage{graphicx}

\begin{document}

\draft \title{Cascading of Liquid Crystal Elastomer Photomechanical Optical Devices} \author{Nathan J. Dawson$^1$, Mark G. Kuzyk$^1$, Jeremy Neal$^2$, Paul Luchette$^2$, and Peter Palffy-Muhoray$^2$}
\address{$^1$Department of Physics and Astronomy, Washington State University, Pullman, WA 99164-2814 \\
$^2$Liquid Crystal Institute, Kent State University, Kent, OH 44242}

\begin{abstract}
Photomechanical actuation is demonstrated in two coupled liquid crystal elastomer photomechanical optical devices (PODs) acting in series. The response function of an individual POD is characterized and used to predict the temporal response of the coupled system. The predicted coupled-system response agrees with the experiment for several waveforms and frequencies, suggesting that large-scale integration of photomechanical devices is possible.
\end{abstract}

\maketitle

\section{Introduction}

\sloppy  Each electronic device class has an optical analogue, i.e. wires/optical fibers, thermistors/interferometric sensors, etc.\cite{nlosource} Having the ability to integrate together all the optical device classes would lead to new devices not possible with electronics, such as smart morphing systems.  Key to such a technology is the availability of mechanically flexible materials that combine logic, sensing, and actuation - as provided with photomechanical materials.  Here, we characterize the response function of an interferometric photomechanical device using a novel liquid crystalline material and show that two cascaded devices behave as predicted by theory for various waveforms and frequencies.

Photomechanical effects have been observed in many materials\cite{uchin93.01,uchin90.01,camac04.01,kuzyk06.06} but the materials with the largest light-induced deformations are side-chain liquid crystal (LC) elastomers.\cite{camac04.01,corbe09.01,Finke01.01,warne05.01} In the mid-1990s, a dye-doped polymer fiber Fabry-Perot cavity was used to make a variety of all-optical photomechanical devices,\cite{welke94.01,welke95.01,welke96.01,welke96.02} including the first all-optical circuit that combined sensing, actuation, and logic.\cite{welke94.01}  Bian and coworkers showed that the dominant mechanisms in a dye-doped polymer are photo-thermal expansion and molecular reorientation.\cite{bian06.01}

\begin{figure}[t]
\includegraphics[scale=1]{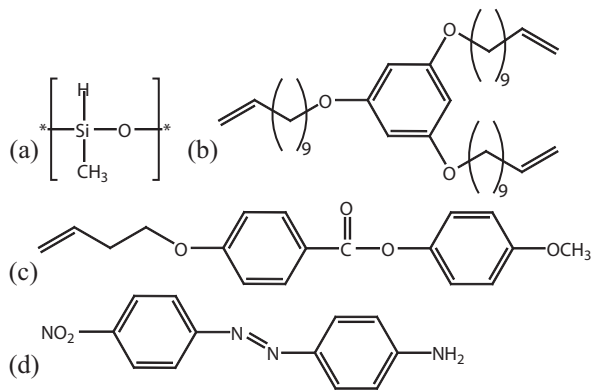}
\caption{(a) The chemical structure of the silicon backbone, (b) tri-functional cross-linker, (c) mesogenic side-chain, and (d) disperse orange 3 azo-dye dopant.}
\label{fig:fourstructure}
\end{figure}

\begin{figure}[b]
\includegraphics[scale=1]{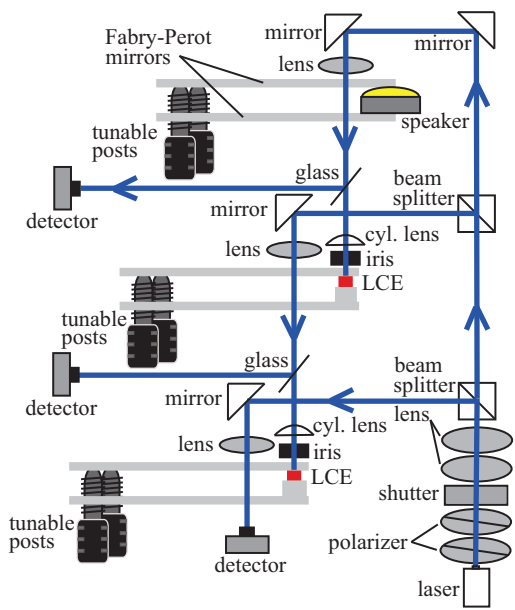}
\caption{The experimental setup for two LCE-interferometer devices in series. A speaker is used inside a Fabry-Perot interferometer to modulate the initial laser intensity.}
\label{fig:setupbw}
\end{figure}

In the present work, we use a highly-dye-doped liquid crystal elastomer (LCE) in a Fabry-Perot interferometer as the basic Photomechanical Optical Device (POD). The molecular structures of the components of a LCE are shown in Figure \ref{fig:fourstructure}.  Integration of PODs in a waveguide could lead to ultra-smart morphing materials with emergent properties.\cite{kuzyk06.06}  Our goal is to demonstrate serial behavior of PODs as a first step in making smart materials.  Figure \ref{fig:setupbw} shows the experiment.  The laser intensity is modulated with a speaker that drives a Fabry-Perot interferometer.  Each device is actuated with one beam and probed by a second beam that passes through an interferometer whose output depends on the length of the LCE.  The probe output from the first elastomer drives the second elastomer.

\section{Experiment}

When the temperature of a liquid crystal increases, the orientational order is reduced.  In an LCE, the liquid crystal and elastomer are coupled so that changes in orientational order of the liquid crystal leads to an internal stress in the elastomer.  The LCEs used in our studies are made by stretching them during the fabrication process, which leads to a bulk material in which the director of the liquid crystals is aligned along the stretching direction.\cite{camac04.01,corbe09.01,Finke01.01,warne05.01}  The LCEs are doped with chromophores that can be optically activated to decrease the liquid crystal's order parameter, thus yielding a net contraction of the bulk material along the LC's director.

Our device takes advantage of the extreme light-induced contraction of an LCE, which is leveraged by the orientationally-ordered nematic phase of the LC.  There are two mechanisms that can lead to a length change: (1) a decrease in the orientational order of the mesogen due to a temperature increase upon absorbing light and (2) a reduction of the liquid crystalline order when the dopant molecule undergoes trans to cis photo-isomerization.\cite{warne05.01}  At room temperature, the thermal contraction coefficient of an LCE\cite{warne05.01} is an order of magnitude larger than the thermal expansion coefficient of a typical thermoplastic polymer.\cite{kuzyk06.06}  The director of the LCE is oriented perpendicular to the partially-reflecting mirror's surface. This contraction of the LCE along the director will decrease the separation between the mirrors that define the interferometer.

The driving laser has a wavelength of $488\mathrm{nm}$, which is on resonance with the Disperse Orange 3 (DO3) dopant.  Energy that is absorbed by the DO3 chromophore is transferred as heat to the LCE to induce thermal contraction or results in photo-isomerization.  The elastomer thickness is much greater than the 1/e absorption length, so the pump beam is fully absorbed.  The probe beam is de-focused prior to entering the interferometer cavity to produce a ring pattern at the exit. The outgoing beam is split into two: one is deflected by a beam splitter to a detector and the other is transmitted as a pump beam to the second LCE. A second beam is used to probe the second interferometer.  The outgoing probe beam is intercepted by a detector.  However, the probe beam exiting the second POD can be used to pump a third stage (not shown), and so on.

Due to the high absorption coefficient of the LCE, most of the light is absorbed near the LCE surface. Therefore, photo-isomerization is also localized near the surface while the photo-generated heat diffuses throughout the LCE. Thus, the response of the POD will depend on these two mechanisms.

The detector measures the center portion of the ring pattern exiting the interferometer.  The Fabry-Perot interferometer rests on two tunable precision screws and the elastomer, forming a tripod support. The LCE is approximately $400\mu\mathrm{m}$ thick and presents a $400\mu\mathrm{m}\times2\mathrm{mm}$ cross-sectional area to the pump beam. The longer length of the LCE is parallel to the line joining the contact points of the two set screws for added stability.

\section{Results and Discussion}

\begin{figure}[b]
\includegraphics[scale=1]{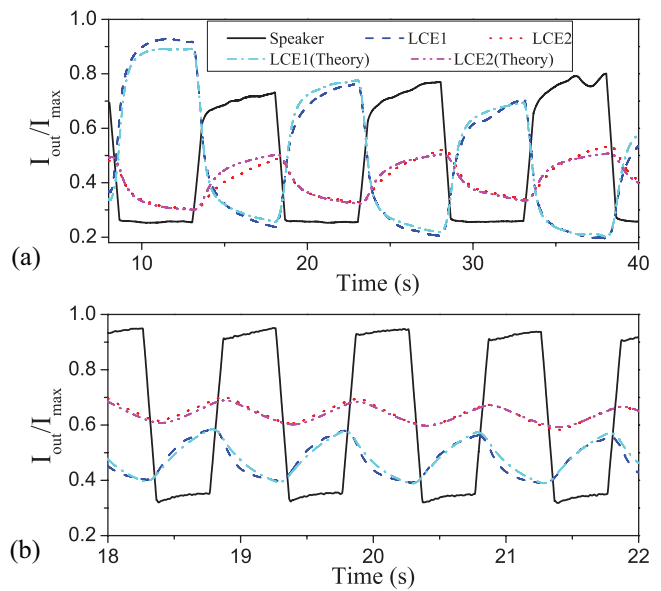}
\caption{The transmitted intensity through each Fabry-Perot interferometer for a square wave input voltage to the speaker interferometer. $I_\mathrm{max}$ is the maximum intensity of the central spot in the interference pattern.  (a) The experimental results and theory for a 0.1Hz waveform (b) and for a 1Hz waveform.}
\label{fig:SQ}
\end{figure}

\begin{figure}[t]
\includegraphics[scale=1]{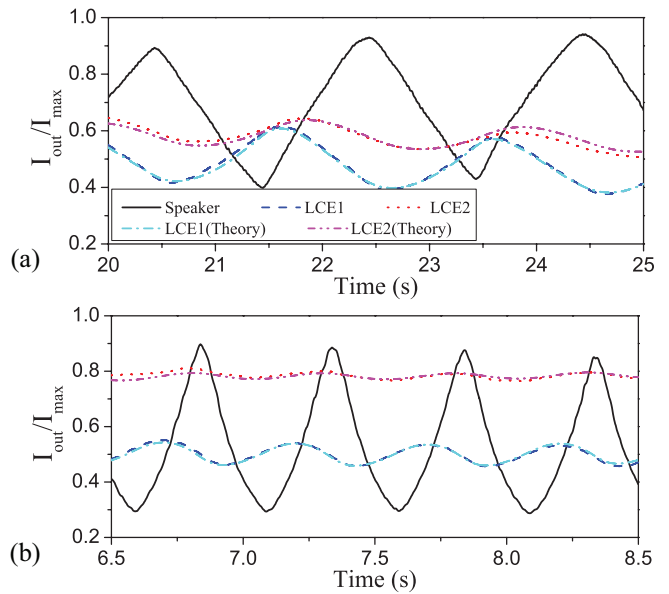}
\caption{(a) The transmitted intensities of each etalon for a 0.5Hz triangular voltage waveform driving the speaker; and, (b) the transmitted intensities for a 2Hz triangle waveform.  $I_\mathrm{max}$ is the maximum intensity of the central spot in the interference pattern.}
\label{fig:TRI}
\end{figure}

A subset of the experimental results are shown in Figures ~\ref{fig:SQ} and ~\ref{fig:TRI}. The small amount of drift in the interferometer is as it is expected from theory for a changing mirror separation due to heating of the substrate and heat diffusion through the interface. Aside from drift due to a slow background temperature drift, the POD's response follows the oscillations of the pump beam.  The response time of the PODs in Figure \ref{fig:SQ}a is faster than the period of the waveform, therefore there is no lag between the waveform and the response.   The $\pi$ phase lag of the first POD is due to the interferometer being tuned to the negative slope part of the interferogram.  The second POD is in phase with the pump because it is tuned to the positive slope region.

Figure ~\ref{fig:TRI}a shows experimental results for a triangular pump intensity waveform whose period is comparable to the POD response time.  There is a clear lag between the waveform and the response of the first POD.  A similar lag is observed between the first and the second POD. Figure ~\ref{fig:TRI}b shows the experimental results for a waveform of a period that exceeds the response time.  In addition to the phase shift of the response, the amplitude of the response of the second POD is lower due the first POD's lower amplitude of light modulation compared with the light source.  We note that because the tuning of each Fabry-Perot interferometer can drift between measurements, some of the differences in phase lag and amplitudes may be due to drift.

The response function of a material may be determined by characterizing the time dependence of the strain.  A phenomena with multiple mechanisms can be modeled by a response function that is the sum of exponentials of the form,
\begin{equation}
R\left(t-\tau\right) = \displaystyle\sum_i^N b_i A_i e^{-b_i \left(t-\tau\right)} ,
\label{eq:respseq}
\end{equation}
where $R$ is the response function of the material, $b_i$ is the time constant and $A_i$ the amplitude of mechanism $i$. The response function relates the input intensity, $I(\tau)$, at time $\tau$ to the strain, $\sigma(t)$, at a later time $t$ according to,
\begin{equation}
\sigma\left(t\right) = \displaystyle\int_{-\infty}^t R\left(t-\tau\right) I\left(\tau\right) \, d\tau .
\label{eq:sigmaint}
\end{equation}

Since the experiment measures the output intensity of the interferometer, $I_\mathrm{out}(t)$,
we must relate the strain to the output intensity using the Airy function,
\begin{equation}
\frac{I_\mathrm{out}\left(t\right)}{I_\mathrm{max}} = \displaystyle\left[1+F\sin^2\left(\frac{2\pi\L_0\sigma\left(t\right)}{\lambda} + \delta\right)\right]^{-1},
\label{eq:fabper}
\end{equation}
where $F$ is the finesse, $L_0$ is the initial length of the LCE, $\lambda$ is the wavelength, and $\delta$ is the phase shift. The sum of three exponentials is sufficient to model the response using Equation \ref{eq:respseq}.  For the one set of fitting parameters shown in Table \ref{table:coeff}, the model yields the theoretical curves over a short range of time (after transients decay and the system reaches equilibrium) as shown is Figures \ref{fig:SQ} and \ref{fig:TRI}.  Only $\delta$ and $I_\mathrm{max}$ needs to be adjusted between runs to account for interferometer drift.

\begin{table}[b]
\caption{LCE Parameters}
\centering
\begin{tabular}{l l l}
\hline\hline
Constant & Value & Units\\ [0.5ex]
\hline
$b_1$ & $5.8 \times 10^{-2}$ & s$^{-1}$\\
$b_2$ & $0.55$ & s$^{-1}$\\
$b_3$ & $3.2$ & s$^{-1}$\\
$A_1$ & $1.23 \times 10^{-8}$ & W$^{-1} \,$m$^2$\\
$A_2$ & $2.30 \times 10^{-9}$ & W$^{-1} \,$m$^2$\\
$A_3$ & $1.66 \times 10^{-8}$ & W$^{-1} \,$m$^2$\\
[1ex]
\hline
\end{tabular}
\label{table:coeff}
\end{table}

The fact that three exponentials provide a good fit to the data suggests that there are no more than 3 mechanisms responsible.  This is consistent with the mechanisms of heating and light-induced reduction of the nematic order of the liquid crystal domains due to photo-isomerization as the two dominant mechanisms.

Relaxation process in polymers often lead to multi-exponential or stretched exponential behavior due to inhomogeneities that are introduced by the distribution of sites.\cite{singe91.01}  We tried to model the response function using a single stretched exponential and found that while fits would be acceptable in limited ranges of the data, no one set of parameters were able to describe all of our data.  Thus, it appears that our material is characterized by a small distinct set of mechanisms rather than an effect due to a continuous distribution of sites.  While the fit to the data may improve if each of the exponentials were to be generalized to stretched exponentials, this seems unnecessary given that the full data set is well characterized by 3 simple exponentials.

\section{Conclusion}

In summary, we have demonstrated the first bulk all-optical cascaded photomechanical optical device using LCEs.  We have characterized the tri-exponential response function of the material/device, and found that all of the data is consistent with the theory that uses {\em one set of parameters}.  The present device is a natural test bed for evaluating new device architectures in the quest for developing an ultra-smart materials technology owing to both the slow and large response of the system.  Miniaturization and integration of such devices into a thermoplastic polymer-based fiber waveguide would increase the device speed and make interactions between PODs more efficient.  Future work will focus on making single-beam fiber Bragg grating cascaded devices.

We acknowledge support of the NSF through grants ECCS-0756936 (MGK) and DMR-0606357 (PPM).

\clearpage

\bibliographystyle{model1a-num-names}

\begin{thebibliography}{13}%
\makeatletter
\providecommand \@ifxundefined [1]{%
 \@ifx{#1\undefined}
}%
\providecommand \@ifnum [1]{%
 \ifnum #1\expandafter \@firstoftwo
 \else \expandafter \@secondoftwo
 \fi
}%
\providecommand \@ifx [1]{%
 \ifx #1\expandafter \@firstoftwo
 \else \expandafter \@secondoftwo
 \fi
}%
\providecommand \natexlab [1]{#1}%
\providecommand \enquote  [1]{``#1''}%
\providecommand \bibnamefont  [1]{#1}%
\providecommand \bibfnamefont [1]{#1}%
\providecommand \citenamefont [1]{#1}%
\providecommand \href@noop [0]{\@secondoftwo}%
\providecommand \href [0]{\begingroup \@sanitize@url \@href}%
\providecommand \@href[1]{\@@startlink{#1}\@@href}%
\providecommand \@@href[1]{\endgroup#1\@@endlink}%
\providecommand \@sanitize@url [0]{\catcode `\\12\catcode `\$12\catcode
  `\&12\catcode `\#12\catcode `\^12\catcode `\_12\catcode `\%12\relax}%
\providecommand \@@startlink[1]{}%
\providecommand \@@endlink[0]{}%
\providecommand \url  [0]{\begingroup\@sanitize@url \@url }%
\providecommand \@url [1]{\endgroup\@href {#1}{\urlprefix }}%
\providecommand \urlprefix  [0]{URL }%
\providecommand \Eprint [0]{\href }%
\@ifxundefined \urlstyle {%
  \providecommand \doi  [0]{\begingroup \@sanitize@url \@doi}%
  \providecommand \@doi [1]{\endgroup \@@startlink {\doibase
  #1}doi:\discretionary {}{}{}#1\@@endlink }%
}{%
  \providecommand \doi  [0]{doi:\discretionary{}{}{}\begingroup
  \urlstyle{rm}\Url }%
}%
\providecommand \doibase [0]{http://dx.doi.org/}%
\providecommand \Doi [0]{\begingroup \@sanitize@url \@Doi }%
\providecommand \@Doi  [1]{\endgroup\@@startlink{\doibase#1}\@@Doi}%
\providecommand \@@Doi [1]{#1\@@endlink}%
\providecommand \selectlanguage [0]{\@gobble}%
\providecommand \bibinfo  [0]{\@secondoftwo}%
\providecommand \bibfield  [0]{\@secondoftwo}%
\providecommand \translation [1]{[#1]}%
\providecommand \BibitemOpen [0]{}%
\providecommand \bibitemStop [0]{}%
\providecommand \bibitemNoStop [0]{.\EOS\space}%
\providecommand \EOS [0]{\spacefactor3000\relax}%
\providecommand \BibitemShut  [1]{\csname bibitem#1\endcsname}%
\bibitem{nlosource}
  see, for example, NLOsource.com
\bibitem [{\citenamefont {Uchino}(1993)}]{uchin93.01}%
  \BibitemOpen
  \bibfield  {author} {\bibinfo {author} {\bibfnamefont {K.}~\bibnamefont
  {Uchino}},\ }\href@noop {} {\bibfield  {journal} {\bibinfo  {journal} {MRS
  Bulletin, April},\ }\textbf {\bibinfo {volume} {29}},\ \bibinfo {pages} {42}
  (\bibinfo {year} {1993})}\BibitemShut {NoStop}%
\bibitem [{\citenamefont {Uchino}(1990)}]{uchin90.01}%
  \BibitemOpen
  \bibfield  {author} {\bibinfo {author} {\bibfnamefont {K.}~\bibnamefont
  {Uchino}},\ }\href@noop {} {\bibfield  {journal} {\bibinfo  {journal}
  {Ultrasonics Symposium},\ \bibinfo {pages} {721}} (\bibinfo {year}
  {1990})}\BibitemShut {NoStop}%
\bibitem [{\citenamefont {Camacho-Lopez}\ \emph {et~al.}(2004)\citenamefont
  {Camacho-Lopez}, \citenamefont {Finkelmann}, \citenamefont {Palffy-Muhoray},\
  and\ \citenamefont {Shelley}}]{camac04.01}%
  \BibitemOpen
  \bibfield  {author} {\bibinfo {author} {\bibfnamefont {M.}~\bibnamefont
  {Camacho-Lopez}}, \bibinfo {author} {\bibfnamefont {H.}~\bibnamefont
  {Finkelmann}}, \bibinfo {author} {\bibfnamefont {P.}~\bibnamefont
  {Palffy-Muhoray}}, \ and\ \bibinfo {author} {\bibfnamefont {M.}~\bibnamefont
  {Shelley}},\ }\href@noop {} {\bibfield  {journal} {\bibinfo  {journal}
  {Nature Materials},\ }\textbf {\bibinfo {volume} {3}},\ \bibinfo {pages}
  {307} (\bibinfo {year} {2004})}\BibitemShut {NoStop}%
\bibitem [{\citenamefont {Kuzyk}(2006)}]{kuzyk06.06}%
  \BibitemOpen
  \bibfield  {author} {\bibinfo {author} {\bibfnamefont {M.~G.}\ \bibnamefont
  {Kuzyk}},\ }\href@noop {} {\emph {\bibinfo {title} {Polymer Fiber Optics:
  materials, physics, and applications}}},\ \bibinfo {series} {Optical science
  and engineering}, Vol.\ \bibinfo {volume} {117}\ (\bibinfo  {publisher} {CRC
  Press},\ \bibinfo {address} {Boca Raton},\ \bibinfo {year}
  {2006})\BibitemShut {NoStop}%
\bibitem [{\citenamefont {Corbett}\ and\ \citenamefont
  {Warner}(2009)}]{corbe09.01}%
  \BibitemOpen
  \bibfield  {author} {\bibinfo {author} {\bibfnamefont {D.}~\bibnamefont
  {Corbett}}\ and\ \bibinfo {author} {\bibfnamefont {M.}~\bibnamefont
  {Warner}},\ }\href@noop {} {\bibfield  {journal} {\bibinfo  {journal} {Liq.
  Cryst.},\ }\textbf {\bibinfo {volume} {36}},\ \bibinfo {pages} {1263}
  (\bibinfo {year} {2009})}\BibitemShut {NoStop}%
\bibitem [{\citenamefont {Finkelmann}\ \emph {et~al.}(2001)\citenamefont
  {Finkelmann}, \citenamefont {Nishikawa}, \citenamefont {Pereira},\ and\
  \citenamefont {Warner}}]{Finke01.01}%
  \BibitemOpen
  \bibfield  {author} {\bibinfo {author} {\bibfnamefont {H.}~\bibnamefont
  {Finkelmann}}, \bibinfo {author} {\bibfnamefont {E.}~\bibnamefont
  {Nishikawa}}, \bibinfo {author} {\bibfnamefont {G.~G.}\ \bibnamefont
  {Pereira}}, \ and\ \bibinfo {author} {\bibfnamefont {M.}~\bibnamefont
  {Warner}},\ }\href@noop {} {\bibfield  {journal} {\bibinfo  {journal} {Phys.
  Rev. Lett.},\ }\textbf {\bibinfo {volume} {87}},\ \bibinfo {pages} {015501}
  (\bibinfo {year} {2001})}\BibitemShut {NoStop}%
\bibitem [{\citenamefont {Warner}\ and\ \citenamefont
  {Terentjev}(2005)}]{warne05.01}%
  \BibitemOpen
  \bibfield  {author} {\bibinfo {author} {\bibfnamefont {M.}~\bibnamefont
  {Warner}}\ and\ \bibinfo {author} {\bibfnamefont {E.~M.}\ \bibnamefont
  {Terentjev}},\ }\href@noop {} {\emph {\bibinfo {title} {Liquid Crystal
  Elastomers}}}\ (\bibinfo  {publisher} {Oxford University Press},\ \bibinfo
  {year} {2005})\BibitemShut {NoStop}%
\bibitem [{\citenamefont {Welker}\ and\ \citenamefont
  {Kuzyk}(1994)}]{welke94.01}%
  \BibitemOpen
  \bibfield  {author} {\bibinfo {author} {\bibfnamefont {D.~J.}\ \bibnamefont
  {Welker}}\ and\ \bibinfo {author} {\bibfnamefont {M.~G.}\ \bibnamefont
  {Kuzyk}},\ }\href@noop {} {\bibfield  {journal} {\bibinfo  {journal} {Appl.
  Phys. Lett.},\ }\textbf {\bibinfo {volume} {64}},\ \bibinfo {pages} {809}
  (\bibinfo {year} {1994})}\BibitemShut {NoStop}%
\bibitem [{\citenamefont {Welker}\ and\ \citenamefont
  {Kuzyk}(1995)}]{welke95.01}%
  \BibitemOpen
  \bibfield  {author} {\bibinfo {author} {\bibfnamefont {D.~J.}\ \bibnamefont
  {Welker}}\ and\ \bibinfo {author} {\bibfnamefont {M.~G.}\ \bibnamefont
  {Kuzyk}},\ }\href@noop {} {\bibfield  {journal} {\bibinfo  {journal} {Appl.
  Phys. Lett.},\ }\textbf {\bibinfo {volume} {66}},\ \bibinfo {pages} {2792}
  (\bibinfo {year} {1995})}\BibitemShut {NoStop}%
\bibitem [{\citenamefont {Welker}\ and\ \citenamefont
  {Kuzyk}(1996){\natexlab{a}}}]{welke96.01}%
  \BibitemOpen
  \bibfield  {author} {\bibinfo {author} {\bibfnamefont {D.~J.}\ \bibnamefont
  {Welker}}\ and\ \bibinfo {author} {\bibfnamefont {M.~G.}\ \bibnamefont
  {Kuzyk}},\ }\href@noop {} {\bibfield  {journal} {\bibinfo  {journal} {Appl.
  Phys. Lett.},\ }\textbf {\bibinfo {volume} {69}},\ \bibinfo {pages} {1835}
  (\bibinfo {year} {1996}{\natexlab{a}})}\BibitemShut {NoStop}%
\bibitem [{\citenamefont {Welker}\ and\ \citenamefont
  {Kuzyk}(1996){\natexlab{b}}}]{welke96.02}%
  \BibitemOpen
  \bibfield  {author} {\bibinfo {author} {\bibfnamefont {D.~J.}\ \bibnamefont
  {Welker}}\ and\ \bibinfo {author} {\bibfnamefont {M.~G.}\ \bibnamefont
  {Kuzyk}},\ }\href@noop {} {\bibfield  {journal} {\bibinfo  {journal} {Nonlin.
  Opt.},\ }\textbf {\bibinfo {volume} {15}},\ \bibinfo {pages} {435} (\bibinfo
  {year} {1996}{\natexlab{b}})}\BibitemShut {NoStop}%
\bibitem [{\citenamefont {Bian}\ \emph {et~al.}(2006)\citenamefont {Bian},
  \citenamefont {Robinson},\ and\ \citenamefont {Kuzyk}}]{bian06.01}%
  \BibitemOpen
  \bibfield  {author} {\bibinfo {author} {\bibfnamefont {S.}~\bibnamefont
  {Bian}}, \bibinfo {author} {\bibfnamefont {D.}~\bibnamefont {Robinson}}, \
  and\ \bibinfo {author} {\bibfnamefont {M.~G.}\ \bibnamefont {Kuzyk}},\
  }\href@noop {} {\bibfield  {journal} {\bibinfo  {journal} {J. Opt. Soc. Am.
  B},\ }\textbf {\bibinfo {volume} {23}},\ \bibinfo {pages} {697} (\bibinfo
  {year} {2006})}\BibitemShut {NoStop}%
\bibitem[{\citenamefont{Singer and King}(1991)}]{singe91.01}
\bibinfo{author}{\bibfnamefont{K.~D.} \bibnamefont{Singer}} \bibnamefont{and}
  \bibinfo{author}{\bibfnamefont{L.~A.} \bibnamefont{King}},
  \bibinfo{journal}{J. Appl. Phys.} \textbf{\bibinfo{volume}{70}},
  \bibinfo{pages}{3251} (\bibinfo{year}{1991}).
\end{thebibliography}

\end{document}